\documentclass[twocolumn,showpacs,preprintnumbers,amsmath,amssymb]{revtex4}
\usepackage{graphicx}
\usepackage{dcolumn}
\usepackage{bm}

\begin{document}

\title{Chaotic instantons and enhancement of tunneling in kicked double-well system with time-reversal symmetry}
\author{V.I. Kuvshinov, A.V. Kuzmin and V.A. Piatrou}
\affiliation{Joint Institute for Power and Nuclear Research,\\ Krasina str. 99, Minsk,  220109, Belarus }
\date{\today}
\begin{abstract}
Chaotic instanton approach is used to describe dynamical tunneling in kicked double well system. Effective Hamiltonian for the kicked system is obtained using matrix expansion formula for operator exponent and exploited to construct an approximation for chaotic instanton solution. This approximation is used for derivation of the ground quasienergy splitting dependence on both the perturbation strength and frequency. Results of numerical calculations for corresponding ground quasienergy splitting dependencies based on Floquet theory are in good agreement with the derived analytical formula in a wide range of perturbation parameters.
\end{abstract}



\maketitle

\section{Introduction}
\addcontentsline{toc}{section}{Introduction}

Investigation of the influence of small perturbation on the behavior of the nonlinear dynamical systems attracts permanent interest for the several last decades  \cite{Liberman, Zaslavski,Haake,Reichl}. The connection between the semiclassical properties of perturbed nonlinear systems and purely quantum processes such as tunneling is a reach rapidly developing field of research nowadays \cite{Haake,Grifoni:98}. Our insight in some novel phenomena in this field was extended during the last decades. The most intriguing among them are the chaos assisted tunneling (CAT) and the closely related coherent destruction of tunneling (CDT).

The former in particular is an enhancement of tunneling in the perturbed low-dimensional systems at small external field strengths and driving frequencies~\cite{Lin:90,Peres:91,Plata:92,Holthaus:92}. This phenomenon takes place when levels of the regular doublet undergo an avoided crossing with the chaotic state~\cite{Bohigas:93,Latka:94}. At the semiclassical level of description one considers tunneling between KAM-tori embedded into the "chaotic sea". The region of chaotic motion affects tunneling rate because compared to direct tunneling between tori it is easier for the system to penetrate primarily into the chaotic region, to travel then along some classically allowed path and finally to tunnel onto another KAM-torus~\cite{Utermann:94, Mouchet:01}. The later, CDT phenomenon, is a suppression of tunneling which occurs due to the exact crossing of two  states with different symmetries from the tunneling doublet \cite{Grossmann:91}. In this case tunneling time diverges which means the total localization of quantum state on the initial torus.

CAT phenomenon as well as CDT were experimentally observed in a number of real physical systems. The CAT observation  between whispering gallery-type modes of microwave cavity having the form of the annular billiard was reported in  Ref.~\cite{Dembowski:00}. The same phenomenon for ultracold atoms was experimentally investigated in Refs.~\cite{Steck:01,Hensinger:01}. The study of the dielectric microcavities provided evidences for CAT as well~\cite{Podolskiy:05}. Both CAT and CDT phenomena were observed in two coupled optical waveguides \cite{vorobeichik:03, Valle:07}. Recently experimental evidence of coherent control of single particle tunneling in strongly driven double well potential was reported in Ref. \cite{kierig:08}.

A number of effective approaches \cite{Scharf:88, Sokolov:00, Rahav:03:pra} are used for analysis of the classical and quantum properties of the perturbed systems~\cite{Brodier:02,Bandyopadhyay:08:ps}. The most common methods which are used to investigate the interplay between semiclassical properties of perturbed nonlinear systems and quantum processes are numerical methods based on Floquet theory~\cite{Utermann:94,Shirley:65,Grifoni:98} and Random Matrix Theory \cite{Leyvraz:96}. Among other approaches we would like to mention the scattering approach for billiard systems~\cite{Doron:95,Frischat:98} and approach based upon the presence of a conspicuous tree structure hidden in a complicated set of tunneling branches~\cite{Shudo:96,Shudo:98,Shudo:01}.

In this paper we will consider the original analytical approach based on instanton technique. Enhancement of tunneling in system with external force in framework of this approach occurs due chaotic instantons which appear in perturbed case. This approach was proposed in \cite{KKS:02,KKS:03,KK:05} and used in~\cite{Igarashi:06}. Chaotic instanton approach will be developed further here using effective model and exploited for description of the enhancement of tunneling in the kicked double well system. The main purpose of the present study is to prove the ability of developed chaotic instanton approach to give quantitative analytical description of tunneling well agreed with independent numerical calculations based on Floquet theory. It will give additional support and pulse for the further development of analytical methods to investigate tunneling phenomenon in quantum systems with mixed classical dynamics. Alternative approach based on quantum instantons which are defined using an introduced notion of quantum action was suggested in~\cite{Jirari:01, Paradis:05}. Analytical approach to describe tunneling in perturbed systems based on nonlinear resonances consideration was developed in \cite{Brodier:02,Brodier:01}. A theory for dynamical tunneling process using fictitious integrable system was recently given in Ref. \cite{Backer:08}.

Double well potential is a special toy model among other nonlinear systems. This system is a simple and nonlinear at the same time. It is convenient to use this model for tunneling analysis. This system is well studied in the nonperturbed case, e.g. on the base of instanton technique~\cite{Polyakov:77,Vainshtein:82} or WKB method~\cite{Zinn-Justin:81}. Double well potential is often used for description of processes which occurred in wide range of real physical systems: such as  flipping of the ammonia molecule \cite{Merzbacher}, transfer of protons along hydrogen bonds in benzoic-acid dimers at low temperatures \cite{Skinner:88,Oppenlander:89} and macroscopic quantum coherence phenomena in superconducting quantum interference devices \cite{Rouse:95,Friedman:96,Friedman:00} and nanomagnets \cite{Awschalom:92,Barco:99}. Perturbation in this paper is regarded in the form of the periodic kicks. One of the attractive features of this type of perturbation is the extensively-investigated simple quantum map which stroboscobically evolves the system from kick $n$ to kick $n+1$. Kicked systems are recently used for experimental realization of a such novel concept as a quantum ratchet \cite{Sadgrove:07,Dana:08}.

The main aim of this study is a development of the chaotic instanton approach on the base of the construction of   model system effectively approximating original chaotic dynamics by some averaged regular trajectories. Tunneling in  kicked double well system is investigated using developed chaotic instanton approach. For this purpose nonperturbed trajectories of the system in imaginary (Euclidean) time are analyzed in section \ref{sec:sys}. Effective model approximating euclidean chaotic dynamics of the original system is constructed using matrix expansion formula for one period evolution operator in section~\ref{sec:effmodel}. Chaotic instanton and its approximation in constructed time-independent  effective model are discussed in section~\ref{sec:ps}. Results obtained by means of the effective model are used in section \ref{sec:formula} to derive analytical formula for lowest  quasienergy doublet splitting dependence on perturbation parameters. Numerical calculations are performed to check the validity of this formula in the section~\ref{sec:num}.

\section{Kicked double-well potential and nonperturbed trajectory action}\label{sec:sys}

Hamiltonian of the particle in the double-well potential can be written down in the following form:
\begin{equation}\label{eq:H}
H_0 = \frac{p^2}{2 m} + a_0\, x^4 - a_2\, x^2,
\end{equation}
where $m$ - mass of the particle, $a_0, a_2$ - parameters of the
potential. We consider the perturbation of the kick-type and choose it as follows:
\begin{equation}\label{eq:V}
V_{per} = \epsilon\, x^2 \sum^{+ \infty}_{n = - \infty} \delta(t- n
T),
\end{equation}
where $\epsilon$ and $T$ are perturbation strength and period, respectively, $t$ - time. Dependence of the perturbation on coordinate was chosen in the form of $x^2$ in order to preserve spatial symmetry in the perturbed system. Hamiltonian of the perturbed system is the following:
\begin{equation}\label{SystemHamiltonian}
H = H_0 + V_{per}.
\end{equation}

We use Euclidean time in this section to work with instantons in section \ref{sec:ps}. Now we implement Wick rotation ($t \rightarrow - i
\tau$) and define Euclidean Hamiltonian in the following form:
\begin{equation}\label{H_E}
\mathcal{H}^E = \mathcal{H}^E_0 - \epsilon\, x^2 \sum^{+ \infty}_{n = - \infty} \delta(\tau- n T),
\end{equation} 
where $\mathcal{H}^E_0$ - nonperturbed Euclidean Hamiltonian which can be written down as follows:
\begin{equation}\label{H0_E}
	\mathcal{H}^E_0 = \frac{p^2}{2 m} - a_0\, x^4 + a_2\, x^2.
\end{equation}

It is seen that all perturbed trajectories, as well as chaotic instantons, are composed of nonperturbed trajectories' parts connected by jumps which are caused by kicks. Thus primarily we will investigate nonperturbed dynamics in this section and obtain the expression for nonperturbed trajectory action near the separatrix. We use the action-angle variables defined in the case of one-dimensional systems. Using standard technique~\cite[section 1.2]{Liberman} for the nonperturbed Hamiltonian  (\ref{H0_E}) we obtain expressions for dependence of these variables on energy~$E$ and coordinate~$x$ in double well system in Euclidean time and write down them in the following form:
\begin{align}\label{action}
J(E) & = \frac{2 a^{3/2}_2 \sqrt{m}}{3 \pi a_0} \sqrt{1 + \sqrt{1 - 4
\frac{a_0}{a^2_2}\, E \,}} \nonumber \\ & \times\left(L(\chi) - \sqrt{1 - 4
\frac{a_0}{a^2_2}\, E \,} \; K(\chi) \right),\\
 \Theta(E, x) & = \frac{\pi}{2 K(\chi)}F\left(\sqrt{\frac{2
a_0}{a_2}}\frac{x}{\sqrt{1 - \sqrt{1 - 4
\frac{a_0}{a^2_2}\, E \,}}}, \chi \right),\notag
\end{align}
where $J$ and $\Theta$ - action and angle variables, respectively,  $K(\chi)$ and $L(\chi)$ - full elliptic integrals of the first and second kinds, respectively, $F(\phi,\chi)$ - elliptic integral of the first kind,  where $\phi$ is integral argument, $\chi$ - its modulus. Modulus $\chi$ can be expressed in the following form:
\[\chi = \frac{1 - \sqrt{1 - 4 \frac{a_0}{a^2_2}\, E
\,}}{1 + \sqrt{1 - 4 \frac{a_0}{a^2_2}\, E \,}}.\]

We expand expression (\ref{action}) in powers of the (Euclidean) energy difference from the separatrix $\xi$  ($\xi=E_{inst}-E$, $E_{inst} = a^2_2/(4 a_0)$ is the energy on the separatrix) to obtain analytical expression for nonperturbed trajectory action (see \cite{KKS:02,KKS:03,KK:03}). For this purpose we use the representation of full elliptic integrals with close to unity modulus $\chi$ \cite[chapter 21.6]{Korn}:
\begin{align}
L(\chi) & = 1 + \frac{1}{2}\left(\Lambda - \frac12\right) \chi' +
\frac{3}{16}\left(\Lambda - \frac{13}{12}\right) {\chi'}^2 + \dots,\notag\\
K(\chi) & = \Lambda + \frac{1}{4}\left(\Lambda - 1\right) \chi' +
\frac{9}{64}\left(\Lambda - \frac{7}{6}\right) {\chi'}^2 + \dots,\notag
\end{align}
where $\Lambda = \ln (4/\sqrt{\chi'})$ and $\chi' = 1 - \chi$.
Substituting these expressions in~(\ref{action}) and neglecting terms higher than linear one we obtain the following estimative relation between the nonperturbed trajectory action and the Euclidean energy difference from the separatrix:
   \begin{equation}\label{eq:S}
   S[x(\tau, \xi)] = \pi J(E_{inst} - \xi) =  S[x_{inst}(\tau, 0)] -
\alpha \, \sqrt{\frac{m}{a_2}} \; \xi,
   \end{equation}
where \(S[x_{inst}(\tau, 0)] = S_{inst} = 2 \sqrt{m} \, a^{3/2}_2 /(3 \, a_0)\) -
nonperturbed instanton action, $\alpha = (1 + 18 \ln 2)/6$ - numerical coefficient.

\section{Effective time-independent model for the kicked system}\label{sec:effmodel}

Now lets construct the effective time-independent Hamiltonian for the double well system with the perturbation of the kick-type. Initially we will calculate this Hamiltonian in the real time. Effective Hamiltonian in Euclidean time will be obtained at the end of the section and it will be used for the Euclidean phase space analysis in the next one. We will construct the effective time-independent Hamiltonian for the system under investigation using the following definition \cite{Scharf:88}:
\begin{equation}\label{def:eff}
	e^{- i \hat{H}_{eff} T} = e^{- i \hat{H}_0 T/2} e^{- i \epsilon \hat{x}^2} e^{- i \hat{H}_0 T/2},
\end{equation} 
where RHS is a one-period evolution operator of the kicked system. During the kicks the influence of the nonperturbed Hamiltonian is negligible in comparison with perturbation. The moment of the kick is chosen in the middle of the period in order to conserve the time-reversal symmetry in the system ($t \rightarrow - t$) \cite{Haake}. We restrict our consideration by sufficiently small values of both the perturbation strength and period ($\epsilon < 0.1$ and $T < 2 \pi/{\omega_0}$, where $\omega_0$ - oscillation frequency near the bottom of the wells). Deriving $H_{eff}$ from definition~(\ref{def:eff}) and using matrix expansion formula for the exponents we can write down effective Hamiltonian in the following way:
\begin{align*}
\hat{H}_{eff} & = \frac{i}{T} \, \ln \left[
1 - i \hat{H}_0 T - i \epsilon \hat{x}^2 - \frac{\hat{H}^2_0 T^2}{2} - \frac{\epsilon^2 \hat{x}^4}{2} \right. \nonumber \\ & -\left.\frac{\epsilon T}{2} (\hat{H}_0 \hat{x}^2 + \hat{x}^2 \hat{H}_0) + O(3)
\right],
\end{align*}
where we show terms up to the second order on $\epsilon$ and $T$. Finally we obtain the effective time-independent Hamiltonian for the kicked system expanding the logarithm in the series:
\begin{align}\label{Heff}
\hat{H}_{eff} &= \hat{H}_0 + \frac{\epsilon \nu}{2 \pi} \hat{x}^2  - \frac{\epsilon T \hbar^2}{6 m} \frac{p^2}{2 m} + \frac{\epsilon T \hbar^2}{3 m} a_0 \hat{x}^4 \nonumber \\ &- \frac{\epsilon T \hbar^2}{6 m} a_2 \hat{x}^2 + \frac{\epsilon^2  \hbar^2}{3 m} \hat{x}^2 + O(3).
\end{align}
Nonperturbed Hamiltonian  (\ref{eq:H}) is the first term. The second term has the same order (zero) as the nonperturbed part on small parameters ($\epsilon, T$). Terms having the first order are absent in the expression (\ref{Heff}). Next terms which are shown in the formula have the second order on perturbation parameters. Terms with higher orders are omitted in the last expression. We restrict our consideration by two first terms and neglect the others. Thus effective Hamiltonian for the kicked double well system describing the tunneling dynamics between wells can be written down in the following form:
\begin{equation}\label{def:hameff}
H_{eff} = \frac{p^2}{2 m} + a_0 x^4 - \left(a_2 - \frac{\epsilon \, \nu}{2 \pi} \right)x^2.
\end{equation}
This Hamiltonian is coincided with nonperturbed Hamiltonian (\ref{eq:H}) when parameter $a_2$ is replaced by $\tilde{a}_2$. The later is defined as follows:
\begin{equation}\label{a2eff}
	\tilde{a}_2 (\epsilon,\nu) = a_2 - \frac{\epsilon  \nu}{2 \pi}.
\end{equation}
In contrast to the kicked system (\ref{SystemHamiltonian}) the effective system with Hamiltonian (\ref{def:hameff}) is time-independent. We should introduce one more restriction for the perturbation parameters variation. This restriction  follows from our assumption that ordinary instanton approach should be valid for the effective potential.
Condition for ordinary instanton approach applicability is the following \cite{Vainshtein:82}:
\begin{equation}\nonumber
S_{eff} = \frac{2 \tilde{a}^{3/2}_2}{3 a_0} \gtrsim 6.
\end{equation}
We obtain the following restriction in the case considered:
\begin{equation}\label{condition}
	\epsilon \, \nu \lesssim (\epsilon \, \nu)_{max} = 2 \pi \left(a_2 - 3 \sqrt[3]{3} \, a^{2/3}_0\right).
\end{equation} 

Thus we should use intermediate values of the perturbation frequency. Frequency should be not small to use matrix expansion formula and not larger than $(\epsilon \, \nu)_{max}/\epsilon$. 

\begin{figure}[!h]
\centering
\includegraphics[angle = 270,width=0.48\textwidth]{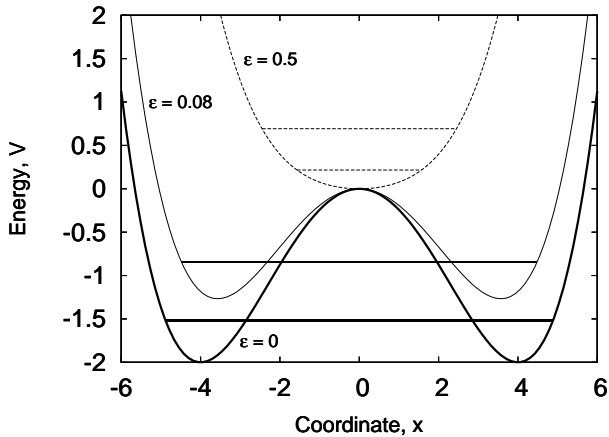}
\caption{Potential and two lowest levels for the effective model. Perturbation frequency $\nu = 4$, perturbation strength $\epsilon = 0$ (thick solid line), $\epsilon = 0.08$ (thin solid line) and $\epsilon = 0.5$ (dashed line). The model parameters are $m~=~1,$ $a_0~=~1/128,$ $a_2 = 1/4$.
}\label{fig:edwp}
\end{figure}

Effective potentials for the set of the perturbation strength and frequency values are shown in the figure~\ref{fig:edwp}. The model parameters for all figures in the paper are the following $m~=~1,$ $a_0~=~1/128,$ $a_2 = 1/4$. The two lowest energy levels in the effective system are shown by horizontal lines in the figure. One can see that for parameters' values satisfying the condition~(\ref{condition}) the two lowest levels are quasidegenerate ($\epsilon = 0$ and $\epsilon = 0.08$). Wells became closer and barrier height is decreased when we increase perturbation parameters. It should enhance tunneling rate in kicked system. These consequences will be checked in following sections for quasienergy spectrum and tunneling process in the system. Two lowest levels are shifted one from another and potential has one well for large values of the perturbation strength  and frequency ($\epsilon = 0.5$ and $\nu =4$ in case shown in the figure~\ref{fig:edwp}). These perturbation parameters are not satisfy the condition (\ref{condition}).

Now we will perform the transformation for effective Hamiltonian (\ref{def:hameff}) into Euclidean time and obtain the following expression:
\begin{equation}\label{def:hameff_E}
H^E_{eff} = \frac{p^2}{2 m} - a_0 x^4 + \left(a_2 - \frac{\epsilon \, \nu}{2 \pi} \right)x^2.
\end{equation}

This Hamiltonian will be used in the following section to analyze the perturbed system phase space in Euclidean time and to construct an approximation for chaotic instanton solution.

\section{Chaotic instantons and effective model Euclidean phase space}\label{sec:ps}

Let us regard a possibility to describe properties of the classical motion in the kicked double well system in Euclidean time using effective model (\ref{def:hameff_E}).We will conduct numerical simulations of the particle dynamics both in  the kicked system and in the framework of effective model for the same set of initial conditions. The deviation between the trajectories in phase space when certain time expires will be a criterion of the applicability of the time-independent approximation in the classical case.

\begin{figure}[!h]
\centering
\includegraphics[angle = 270,width=0.48\textwidth]{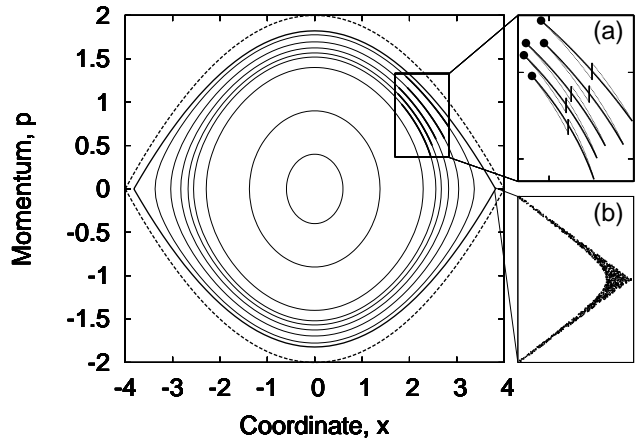}
\caption{Effective model phase space for the kicked system with perturbation parameters $\epsilon = 0.02$ and $\nu = 7$. Separatrix in the effective model (thick solid line) and in the nonperturbed system (dashed line) are shown in the figure. Comparison of the particle classical motion on one period of the perturbation in effective  (thick solid lines) and kicked (thin solid lines) systems from the set of initial conditions (thick points) are shown in the inset $(a)$. Inset $(b)$ shows chaotic trajectory of the kicked particle near the turning point of the effective separatrix.}\label{fig:effps}
\end{figure}

We will consider kicked system (\ref{H_E}) with the perturbation parameters $\epsilon = 0.02$ and $\nu = 7$ and the effective model with parameter $\tilde{a}_2$ which is defined by the expression~(\ref{a2eff}). Separatrix in the effective model is shown in the figure~\ref{fig:effps} by thick solid line while separatrix in nonperturbed system by dashed line. Effective trajectories with energies which are less then on the separatrix are shown by thin solid lines. Comparison of the classical motion on one period of the perturbation in the effective  (thick solid lines) and kicked (thin solid lines) systems for the set of initial conditions (thick points) are shown in the inset $(a)$  in the figure~\ref{fig:effps}. Trajectory in the effective model is a smooth line in the phase space. The trajectory for the perturbed system is a broken line due to the kick in the middle of the period. Trajectories for two systems lie close to each other as it is shown in the inset $(a)$ in the figure~\ref{fig:effps}. Inset~$(b)$ shows chaotic trajectory of the kicked particle near the turning point of the effective separatrix.

In order to check applicability of the effective model for the description of the perturbed system phase space we perform numerical simulations of the particle dynamics using direct simulations for the kicked system and  calculations using effective Hamiltonian (\ref{def:hameff_E}) from the set of initial points. Energy range for this set was chosen from the bottom of the wells ($E_{bottom} = 0$) to the effective model separatrix energy which can be written down in the following way:
\begin{equation}\label{E_top}
	E_{top} = \tilde{a}_2(\epsilon,\nu)^2/(4 a_0).
\end{equation}
Fifty energy values were chosen in this energy interval. One hundred of initial points were randomly evaluated for each energy value. Effective model sufficiently strictly describe the classical dynamics of particles in the kicked double well system for all particle energy range. Mean arithmetical deviation in energy is less then $0.1\%$ of the barrier height. Thus phase portrait of the effective model coincides with the stroboscopic map for the kicked system besides separatrix region. Narrow stochastic layer should be placed near the separatrix in the kicked system. Inset~$(b)$ shows this layer near the turning point of the effective separatrix.

Euclidean equations of motion of the particle in the nonperturbed double-well potential ($\epsilon = 0$) have a well known solution - instanton. This solution is used for calculation of the ground energy splitting in the system without perturbation \cite{Polyakov:77,Vainshtein:82} and explains the rate of the tunneling process in it. Another solutions of the Euclidean equations of motion besides ordinary instanton are required to explain dynamical tunneling in perturbed system. Perturbation destroys the separatrix and some trajectories in its vicinity go to infinity. Narrow stochastic layer is formed nearby the nonperturbed separatrix due to the perturbation. ``Chaotic instanton'' is appeared in this layer. It consists of nonperturbed trajectory's parts connected by jumps. Chaotic instanton is the closest to the destroyed separatrix perturbed trapped trajectory. Thus it plays a dominant role in tunneling in perturbed system. This configuration will be investigated analytically in this section.

Chaotic instanton from dynamical point of view is a set of nonperturbed trajectory parts connected by jumps. These jumps are induced by kicks with period $T$. Energy of the nonperturbed trajectories will be changed after a kick. Effective model can be used for analysis of the kicked system phase space. Thus we can use effective model separatrix as an approximation for the chaotic instanton in the kicked system. Using this approximation we can write down an expression for description of chaotic instanton in phase space:
\begin{equation}\label{momentum}
 p_s(x) = \sqrt{2m \left(\frac{\tilde{a}_2 (\epsilon,\nu)^2}{4 a_0} + a_0 x^4 - \tilde{a_2}
x^2\right)}.
\end{equation}
This approximation, i.e. effective separatrix, is shown by thick solid line in the figure~\ref{fig:effps}.
Energy of the particle which is moving over the chaotic instanton trajectory is changing from maximum energy when particle momentum is zero to minimum at point where particle coordinate is zero.

Lets write down expressions for these energy borders in explicit form. Chaotic instanton energy is maximum for  $p_s(x_{\pm}) = 0$. This point is a turning point for the effective model ($x_{\pm} = \pm \sqrt{\tilde{a}_2/(2 a_0)}$). Using condition for turning points we can write down expression for maximum energy in the following form:
\begin{align}\label{Emax}
	E_{max} &= \mathcal{H}^E_0(p_s=0,x_{\pm}) = \frac{a^2_2}{4 a_0} - \frac{(\epsilon \nu)^2}{16 a_0 \pi^2} \nonumber \\ &= E_{inst} - \frac{(\epsilon \nu)^2}{16 a_0 \pi^2} \approx E_{inst},
\end{align} 
where $E_{inst} = {a^2_2}/({4 a_0})$ - energy of nonperturbed instanton. We restrict our consideration by small values of the perturbation strength $\epsilon$ and intermediate values of the perturbation frequency~$\nu$. Thus we neglect the second term in the expression~(\ref{Emax}). 

Chaotic instanton minimal energy is calculated from the condition $x = 0$ for expression (\ref{momentum}). This energy coincides with the energy on the effective model separatrix (\ref{E_top}). Minimal energy can be written down in the following way:
\begin{align}\label{Emin}
	E_{min} &= \mathcal{H}^E_0(p_s(0),x =0) =  E_{top}  = \tilde{a}_2 (\epsilon,\nu)^2/(4 a_0) \nonumber \\ &= \frac{a^2_2}{4 a_0} - \frac{a_2 \epsilon \nu}{4 a_0 \pi} + \frac{(\epsilon \nu)^2}{16 a_0 \pi^2} \approx E_{inst} - \frac{a_2 \epsilon \nu}{4 a_0 \pi},
\end{align} 
where we have neglected terms higher than the linear one.

The nonperturbed trajectory with energy $E_{max}$ is close to the nonperturbed separatrix. The nonperturbed trajectory with energy $E_{min}$ coincides with the effective separatrix for point $x = 0$ and $p_s(0)$.  Using the last two expressions (\ref{Emax}) and (\ref{Emin}) we obtain the formula for energy range of the chaotic instanton:
\begin{equation}\label{eq:dH}
	\Delta \mathcal{H}^E_{ch.inst.} = E_{max}  - E_{min} = \frac{a_2\,
\epsilon \, \nu}{4 \, a_0 \, \pi}.
\end{equation} 
Expression (\ref{eq:dH}) will be used in the following section in order to obtain analytical formula for the lowest  quasienergy doublet splitting dependence on the perturbation parameters in the kicked system.

\section{Ground doublet quasienergy splitting formula}\label{sec:formula}

The lowest doublet energy splitting in two loop approximation in the nonperturbed double well potential is the following (\cite{Wohler:94} and review \cite{Vainshtein:82}):
\begin{equation}\label{dE_0}
 \Delta E_0 = 2 \, \omega_0 \sqrt{\frac{6}{\pi}} \, \sqrt{S_{inst}} \, exp\left(- S_{inst} - \frac{71}{72}
\frac{1}{S_{inst}}\right),
\end{equation}
where $\omega_0$ - oscillation frequency near the bottom of the wells , $S_{inst}$ - nonperturbed instanton action.

Ground doublet quasienergy splitting ($\Delta \eta$) in the kicked system in the framework of our approach is expressed in terms of chaotic instanton action ($S_{ch}$) through the formula which is similarly to (\ref{dE_0}):
\begin{equation}\label{eq:dn}
	\Delta \eta = 2 \, \omega_0 \sqrt{\frac{6}{\pi}} \, \sqrt{S_{ch}} \, exp\left(- S_{ch} - \frac{71}{72}
\frac{1}{S_{ch}}\right),
\end{equation}
where chaotic instanton action can be calculated by averaging the nonperturbed trajectory action~(\ref{eq:S}) over energy from minimum to maximum for the chaotic instanton energy ($E_{min}$ and $E_{max}$, respectively):
\[S_{ch} = \frac{1}{\Delta \mathcal{H}^E_{ch.inst.}}
\,\int^{E_{max}}_{E_{min}}  S(E)  d\,E.\]
This integral can be transformed to the integral over the energy difference $\xi=E_{inst}-E$ and calculated directly. As a result we have the following expression:
\begin{align}\label{eq:Sch}
S_{ch} &= \frac{1}{\Delta \mathcal{H}^E_{ch.inst.}}
\,\int^{\Delta \mathcal{H}^E_{ch.inst.}}_{0}  S(\xi)  d\,\xi \nonumber \\ &= S_0 - \frac{\alpha}{2} \sqrt{\frac{m}{a_2}}\; \Delta
\mathcal{H}^E_{ch.inst.}.
\end{align}

Now we can write down analytical formula for the ground quasienergy levels splitting using expressions~(\ref{eq:dH}), (\ref{dE_0}), (\ref{eq:dn}) and (\ref{eq:Sch}):
\begin{equation}\label{an-fomula}
\Delta \eta(\epsilon, \nu) = \Delta E_0 \, e^{k \, \epsilon \, \nu},
\end{equation}
where $$k = \frac{\alpha \, \sqrt{m \, a_2}}{8 \, \pi \, a_0}.$$

Tunneling period in the kicked double well potential is expresses in terms of ground quasienergy levels splitting through the following formula:
\begin{equation}\label{eq:T}
	T_{tun} = \frac{2\, \pi}{\Delta\, \eta}.
\end{equation}
Increasing of the perturbation parameters gives exponential rise to ground quasienergy splitting and to the tunneling frequency ($\nu_{tun} (\epsilon, \nu) = \Delta\, \eta (\epsilon, \nu)$). The last exponential factor in the expression~(\ref{an-fomula}) is responsible for the tunneling enhancement in the perturbed system. In nonperturbed case formula (\ref{an-fomula}) coincides with the expression (\ref{dE_0}). Formulas~(\ref{an-fomula}) and~(\ref{eq:T}) will be checked in numerical calculations in the next section.

\section{Numerical calculations}\label{sec:num}

For the computational purposes it is convenient to choose the eigenvectors of harmonic oscillator as the basis vectors. In this representation matrices of the Hamiltonian (\ref{eq:H}) and the perturbation~(\ref{eq:V}) are real and symmetric. They have the following forms ($n \ge m$):
\begin{align*}
H^0_{m\, n} &= \delta_{m \;n} \left[\hbar \omega \left(n + \frac12\right) + \frac g 2 \left(\frac32 \, g\, a_0 \, (2 m^2 + 2  m + 1)\right. \right.\\ & -\left.\left. a'_2 (2 m + 1) \right)  \right] \\  
&+ \delta_{m + 2 \; n} \;\frac{g}{2} \left(g\, a_0  (2 m + 3) - a'_2 \right) \sqrt{(m + 1)(m  + 2)}\\
&+ \delta_{m + 4 \; n} \frac{a_0 g^2}{4} \sqrt{(m + 1)(m + 2)(m + 3)(m + 4)},\\
V_{m\, n} &= \epsilon \; \frac{g}{2} \; \left(\delta_{m + 2 \; n}\; \sqrt{(m + 1)(m  + 2)} +  \delta_{m \;n} (2 m + 1)\right),
\end{align*}
where $g  = \hbar/m \omega$ and $a'_2 = a_2 + m \,\omega^2/2$, $\hbar$ is Planck constant which we put equal to $1$, $\omega$ - frequency of the basis harmonic oscillator which is arbitrary, and so may be adjusted to optimize the computation. We use the value $\omega = 0.2$ with parameters $m~=~1,$ $a_0~=~1/128,$ $a_2 = 1/4$ which are chosen in such a way that nonperturbed instanton action is large enough for energy splitting formula for nonperturbed system to be valid and not too big in order to decrease errors of numerical calculations. The matrix size is chosen to be equal to $200 \times 200$. Calculations with larger matrices give the same results. System of computer algebra Mathematica was used for numerical calculations.

\begin{figure}[!h]
\centering
\includegraphics[angle = 270,width=0.48\textwidth]{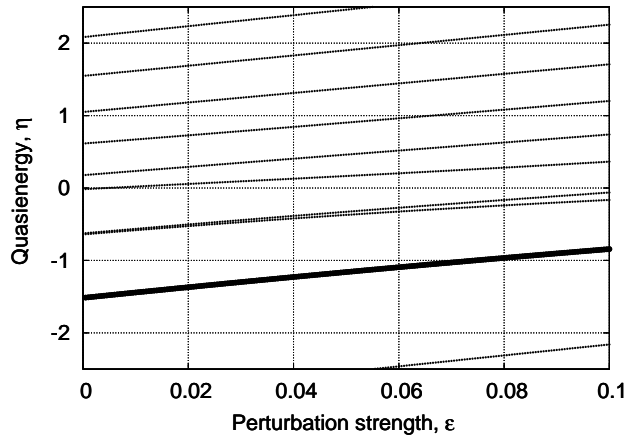}
\caption{Quasienergy spectrum for the ten lowest average energy levels. Thick lines - doublet with the minimal average energy. Perturbation frequency $\nu = 5$.}\label{fig:qeS}
\end{figure}

We calculate eigenvalues of the one-period evolution operator (RHS of the expression (\ref{def:eff})) and obtain quasienergy levels ($\eta_k$) which are related with the evolution operator eigenvalues ($\lambda_k$) through the expression $\eta_k = i \, \ln \lambda_k/T$. Then we get ten levels with the lowest one-period average energy which are calculated using the formula $\left<v_i\right|H_0 + V/T\left|v_i\right>$ ($\left|v_i\right>$ are the eigenvectors of the one-period evolution operator). The dependence of ten lowest levels quasienergies on the strength of the perturbation is shown in the figure~\ref{fig:qeS}. Quasienergies of two levels with the minimal average energy are shown by thick lines in the figure~\ref{fig:qeS}. They are too close to each other to be resolved in the figure due to very small splitting.

\begin{figure}[!ht]
\centering
\includegraphics[angle = 270,width=0.48\textwidth]{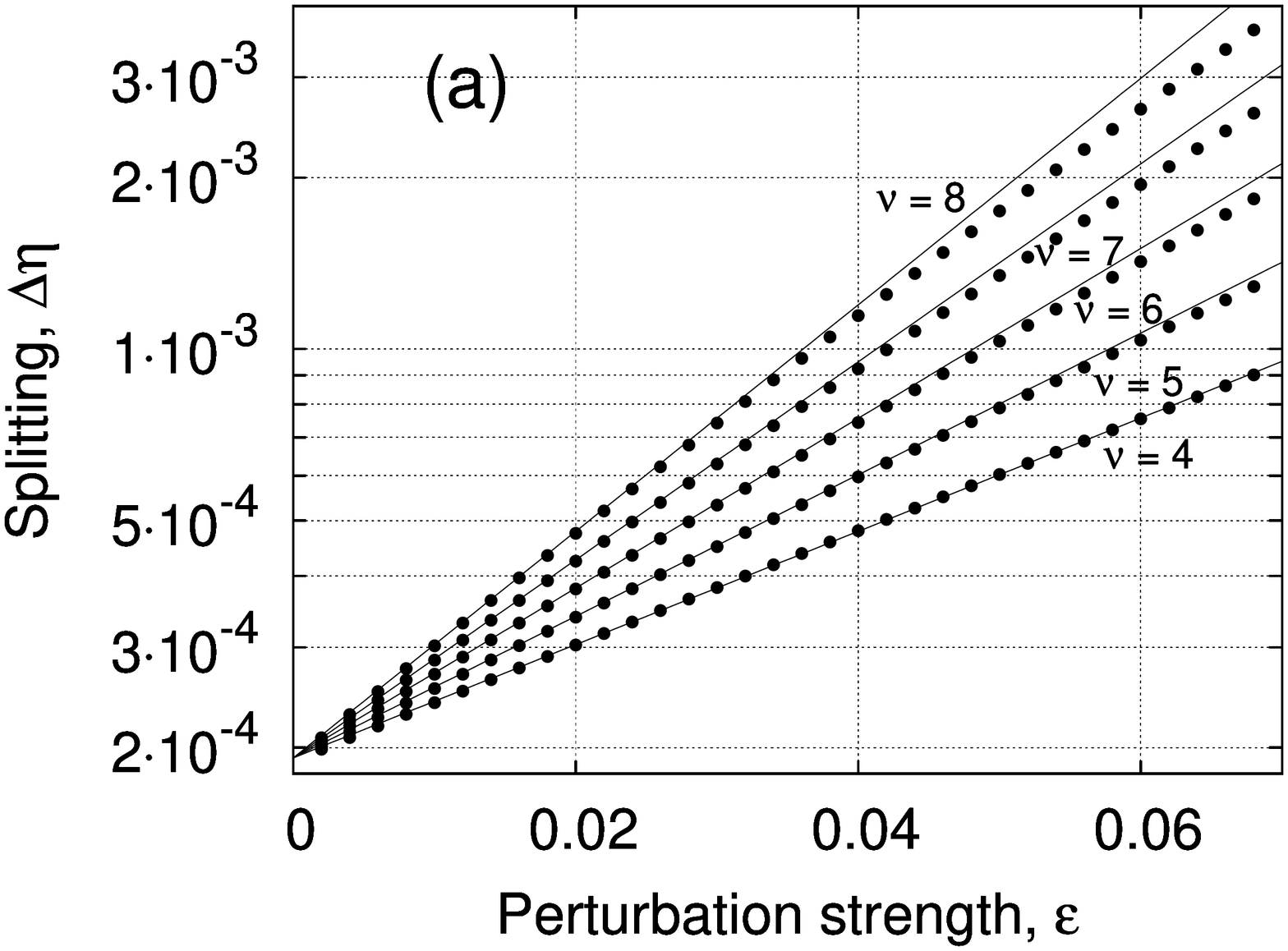}
\includegraphics[angle = 270,width=0.48\textwidth]{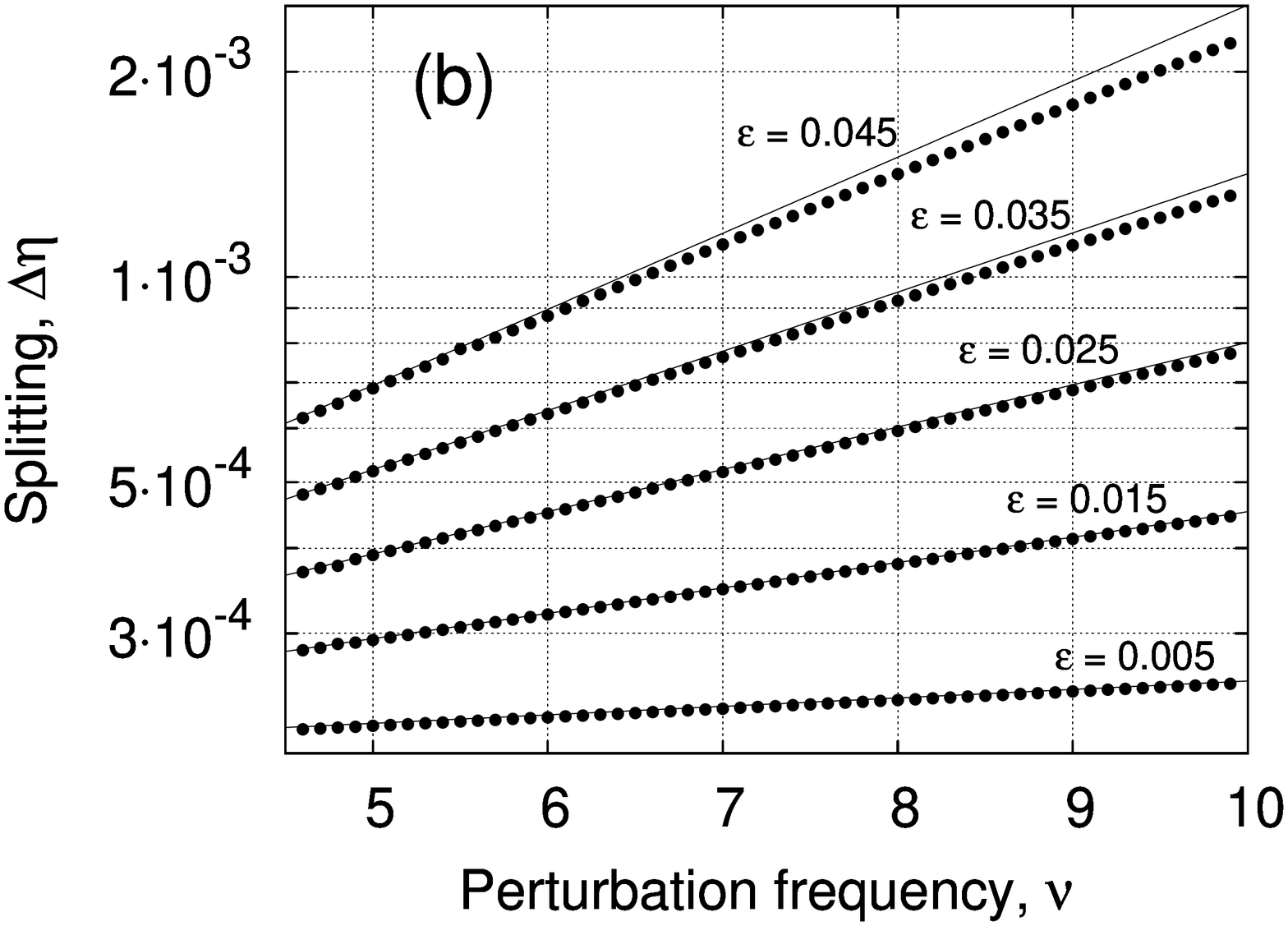}
\caption{Quasienergy splitting as a function of the strength~($a$) and
frequency~($b$) of the perturbation. Lines - analytical
formula (\ref{an-fomula}), points - numerical results.}\label{fig:dE}
\end{figure}

Performed numerical calculations give the dependence of the ground quasienergy splitting both on the strength (fig.\ref{fig:dE}(a)) and the frequency (fig.\ref{fig:dE}(b)) of the perturbation. Results of numerical calculations are plotted in the figure~\ref{fig:dE} by points. Axis $\Delta\eta$ is shown in logarithmic scale. Obtained dependencies are exponential as it was predicted by chaotic instanton approach and obtained analytical formula (\ref{an-fomula}).

Analytical results are plotted in the figures~\ref{fig:dE}~(a)~and~(b) by straight solid lines. Numerical points lie close to these lines. The agreement between numerical calculations and analytical expression is good in the parametric region considered.

\begin{figure}[!ht]
\centering
\includegraphics[angle = 270,width=0.48\textwidth]{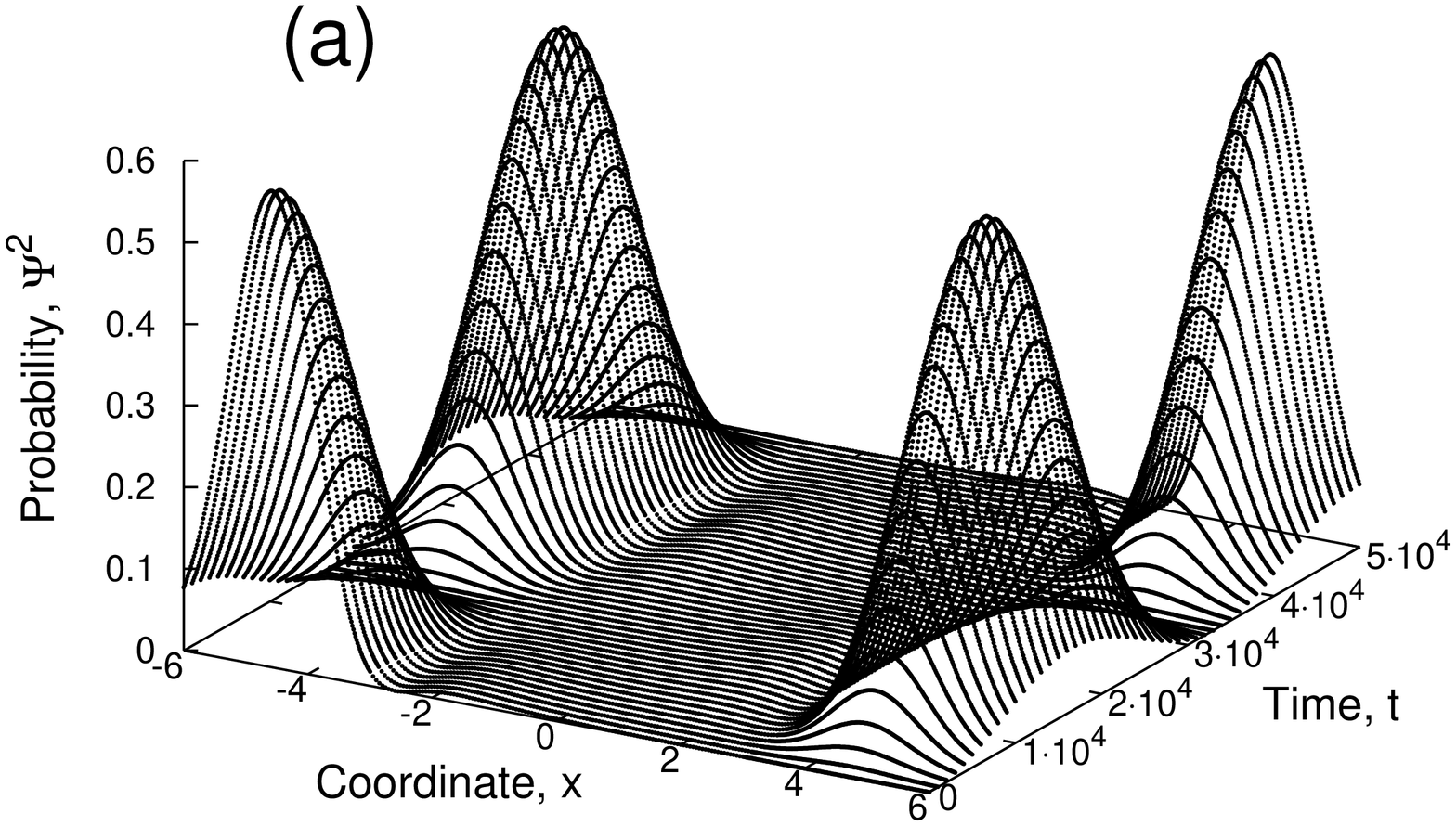}\hfil
\includegraphics[angle = 270,width=0.48\textwidth]{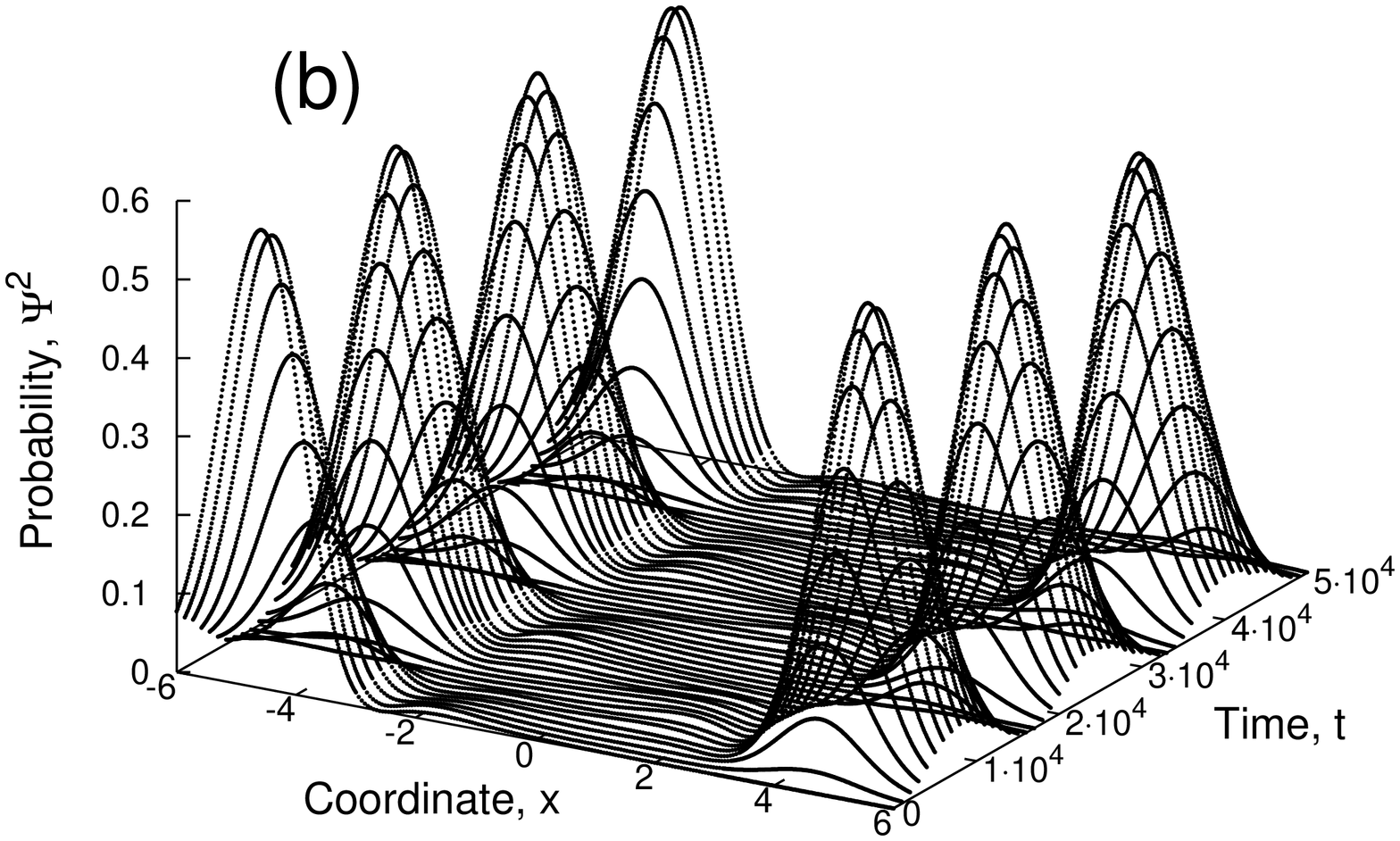}\\
\caption{Quantum mechanical tunneling in kicked double well potential. Perturbation parameters: $\nu = 4$, $(a) \epsilon = 0$, $(b) \epsilon = 3 \cdot 10^{-2}$.}\label{fig:tun}
\end{figure}

Now lets perform numerical simulations for the tunneling process in the kicked double well system and check an applicability of the formulas (\ref{an-fomula}) and (\ref{eq:T}) for this process. For this purpose we regard the double well potential (\ref{eq:H}) with the same parameters and basis vectors as for previous calculations. We take a symmetric superposition of two lowest nonperturbed states as a initial wave packet. These packet is localized in left well of potential. Numerical simulations we provide by multiplying initial wave function by one period evolution operator $e^{- i \hat{H}_0 T/2} e^{- i \epsilon \hat{x}^2} e^{- i \hat{H}_0 T/2}$. The results of numerical simulations for the two sets of the perturbation parameters are shown in the figure~\ref{fig:tun}. The dependence of the localization probability of the wave packet on the coordinate and time is presented in figures. Minima of the nonperturbed double well potential (\ref{eq:H}) are situated in points $x = -4$ and $x = 4$. Tunneling between these points in nonperturbed system is demonstrated in the figure~\ref{fig:tun}(a). Evolution of the initial wave packet in perturbed case is shown in the figure~\ref{fig:tun}(b). Perturbation parameters for these simulations are $\nu = 4$ and $\epsilon = 3 \cdot 10^{-2}$. They are chosen in such a way to speed up a tunneling in two times in comparison with nonperturbed system. Figures \ref{fig:tun}(a) and \ref{fig:tun}(b) demonstrate this enhancement. Fourier analysis of the dependence of the localization probability of the wave packet in left well on time in perturbed case confirms analytical assumptions mentioned above.

\begin{figure}[!h]
\centering
\includegraphics[angle = 270,width=0.48\textwidth]{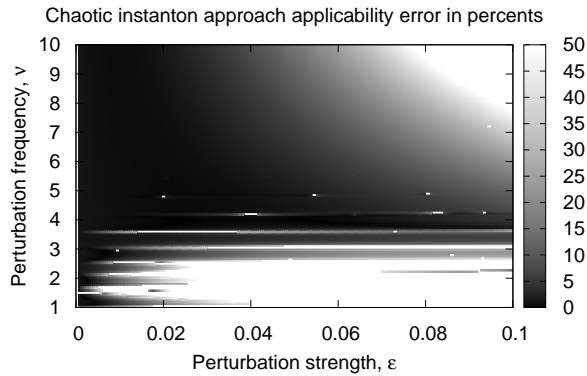}
\caption{Comparison error of the analytical formula~(\ref{an-fomula}) with results of numerical calculations in percents.}\label{fig:appl}
\end{figure}

In order to check applicability of the developed approach we carry out a series of the numerical calculations for wide range of the perturbation parameters. Result of the analysis is performed in figure~\ref{fig:appl}. Region of quantitative agreement between analytical and numerical results is shown by black color in the figure. There are two restrictions of the developed approach applicability. The first one is that the perturbation frequencies have to be not too small. Otherwise the chaotic instanton approach is broken due to restriction for value of perturbation period ($\epsilon < 0.1$ and $T < 2 \pi/{\omega_0}$) which was introduced in section \ref{sec:effmodel}. Another restriction for analytical predictions is a condition which imply the limit for product of perturbation strength and frequency (see \ref{condition}). These two restrictions explain accurately the figure ~\ref{fig:appl}.

Inverse sign in the expression of the perturbation (\ref{eq:V}) will induce exponential suppression of tunneling in the system. It can be demonstrated numerically as well.

\section{Conclusions}
\addcontentsline{toc}{section}{Conclusions}

Chaotic instanton approach allows to describe analytically the influence of the perturbation on quantum properties of nonlinear systems. The kicked double well system is regarded as a toy model to compare quantitative analytical predictions with the results of numerical calculations.

Chaotic instanton is the solution of the Euclidean equations of motion of the perturbed system. This configuration is responsible for the enhancement of tunneling far away from the exact (avoided) level crossings. Time-independent effective model is used for regular approximation of the chaotic instanton solution in order to take into account it's contribution to the ground quasienergy doublet splitting. The effective Hamiltonian is constructed for the kicked double well system using matrix expansion formula for operator exponent for the small values of the strength and period of the perturbation. Dynamical simulations for the perturbed system and effective model show that effective Hamiltonian can be used for the description of the perturbed system Euclidean phase space. The chaotic instanton approximation was constructed and exploited to obtain the energy range for the chaotic instanton trajectory.
Formula for ground quasienergy levels splitting was evaluated  averaging trajectory action in the obtained energy range in the framework of chaotic instanton approach. This formula predicts exponential dependence of the ground doublet splitting on both the perturbation strength and frequency. 

Numerical calculations for quasienergy levels dependence on the perturbation parameters and simulations for tunneling dynamics are performed to check the validity of the obtained analytical formula. Results of numerical calculations for the quasienergy spectrum  confirm the exponential dependence of the ground splitting on both the perturbation strength and frequency. They are in good agreement with the derived analytical formula~(\ref{an-fomula}). 
Simulations of the tunneling dynamics of the wave packet in the kicked double well system demonstrate exponential tunneling enhancement as well. Applicability of chaotic instanton approach was tested in a series of numerical calculations. Sufficiently wide range of perturbation parameters was found suitable for developed approach application.


\begin{thebibliography}{10}

\bibitem{Liberman}
A.~J. Lichtenberg and M.~A. Liberman,
\newblock {\em Regular and Chaotic Dynamics} (Springer-Verlag, New York, 1992).

\bibitem{Zaslavski}
R.~Z. Sagdeev, D.~A. Ousikov, and G.~M. Zaslavski,
\newblock {\em Nonlinear physics: from the pendulum to turbulence and chaos}
  (Harwood Academic Pub (Chur [Switzerland] ; Philadelphia), 1988).

\bibitem{Haake}
F.~Haake,
\newblock {\em Quantum signatures of chaos}, second ed. (Springer-Verlag,
  Berlin, Heidelberg, 2001).

\bibitem{Reichl}
L.~E. Reichl,
\newblock {\em The transition to chaos}, second ed. (Springer-Verlag, 2004).

\bibitem{Grifoni:98}
M.~Grifoni and P.~Hanggi,
\newblock Phys. Rep. {\bf 304}, 229—354 (1998).

\bibitem{Lin:90}
W.~A. Lin and L.~E. Ballentine,
\newblock Phys. Rev. Lett. {\bf 65}, 2927 (1990).

\bibitem{Peres:91}
A.~Peres,
\newblock Phys. Rev. Lett. {\bf 67}, 158 (1991).

\bibitem{Plata:92}
J.~Plata and J.~M.~G. Llorente,
\newblock J. Phys. A: Math. Gen. {\bf 25}, L303 (1992).

\bibitem{Holthaus:92}
M.~Holthaus,
\newblock Phys. Rev. Lett. {\bf 69}, 1596 (1992).

\bibitem{Bohigas:93}
O.~Bohigas, S.~Tomsovic, and D.~Ullmo,
\newblock Phys. Rep. {\bf 223}, 43 (1993).

\bibitem{Latka:94}
M.~Latka, P.~Grigolini, and B.~J. West,
\newblock Phys. Rev. A {\bf 50}, 1071 (1994).

\bibitem{Utermann:94}
R.~Utermann, T.~Dittrich, and P.~H\"anggi,
\newblock Phys. Rev. E {\bf 49}, 273 (1994), arXiv:chao-dyn/9310006v1.

\bibitem{Mouchet:01}
A.~Mouchet, C.~Miniatura, R.~Kaiser, B.~Gremaud, and D.~Delande,
\newblock Phys. Rev. E {\bf 64}, 016221 (2001), arXiv:nlin.CD/0012013v1.

\bibitem{Grossmann:91}
F.~Grossmann, T.~Dittrich, P.~Jung, and P.~H\"anggi,
\newblock Phys. Rev. Lett. {\bf 67}, 516 (1991).

\bibitem{Dembowski:00}
C.~Dembowski {\em et~al.},
\newblock Phys. Rev. Lett. {\bf 84}, 867 (2000), arXiv:chao-dyn/9911023v2.

\bibitem{Steck:01}
D.~A. Steck, W.~H. Oskay, and M.~G. Raizen,
\newblock Science {\bf 293}, 274 (2001).

\bibitem{Hensinger:01}
W.~K. Hensinger {\em et~al.},
\newblock Nature {\bf 412}, 52 (2001).

\bibitem{Podolskiy:05}
V.~A. Podolskiy and E.~E. Narimanov,
\newblock Optics Letters {\bf 30}, 474 (2005).

\bibitem{vorobeichik:03}
I.~Vorobeichik, E.~Narevicius, G.~Rosenblum, M.~Orenstein, and N.~Moiseyev,
\newblock Phys. Rev. Lett. {\bf 90}, 176806 (2003).

\bibitem{Valle:07}
G.~D. Valle {\em et~al.},
\newblock Phys. Rev. Lett. {\bf 98}, 263601 (2007), arXiv:quant-ph/0701121v1.

\bibitem{kierig:08}
E.~Kierig, U.~Schnorrberger, A.~Schietinger, J.~Tomkovic, and M.~K. Oberthaler,
\newblock Phys. Rev. Lett. {\bf 100}, 190405 (2008), arXiv:0803.1406v1
  (quant-ph).

\bibitem{Scharf:88}
R.~Scharf,
\newblock J. Phys. A: Math. Gen. {\bf 21}, 2007 (1988).

\bibitem{Sokolov:00}
V.~V. Sokolov, O.~V. Zhirov, D.~Alonso, and G.~Casati,
\newblock Phys. Rev. Lett. {\bf 84}, 3566 (2000).

\bibitem{Rahav:03:pra}
S.~Rahav, I.~Gilary, and S.~Fishman,
\newblock Phys. Rev. A {\bf 68}, 013820 (2003), arXiv:nlin/0301033v2.

\bibitem{Brodier:02}
O.~Brodier, P.~Schlagheck, and D.~Ullmo,
\newblock Ann. Phys. {\bf 300}, 88 (2002), arXiv:nlin/0205054v1.

\bibitem{Bandyopadhyay:08:ps}
M.~Bandyopadhyay,
\newblock Phys. Scr. {\bf 77}, 055006 (7pp) (2008).

\bibitem{Shirley:65}
J.~H. Shirley,
\newblock Phys. Rev. {\bf 138}, B979 (1965).

\bibitem{Leyvraz:96}
F.~Leyvraz and D.~Ullmo,
\newblock J. Phys. A {\bf 29}, 2529 (1996), arXiv:chao-dyn/9510005v2.

\bibitem{Doron:95}
E.~Doron and S.~D. Frischat,
\newblock Phys. Rev. Lett. {\bf 75}, 3661 (1995), arXiv:cond-mat/9505010v3.

\bibitem{Frischat:98}
S.~D. Frischat and E.~Doron,
\newblock Phys. Rev. E {\bf 57}, 1421 (1998), arXiv:chao-dyn/9707005.

\bibitem{Shudo:96}
A.~Shudo and K.~S. Ikeda,
\newblock Phys. Rev. Lett. {\bf 76}, 4151 (1996).

\bibitem{Shudo:98}
A.~Shudo and S.~Kensuke,
\newblock Physica D: Nonlinear Phenomena {\bf 115}, 234 (1998).

\bibitem{Shudo:01}
T.~Onishi, A.~Shudo, K.~S. Ikeda, and K.~Takahashi,
\newblock Phys. Rev. E {\bf 64}, 025201 (2001), arXiv:nlin/0105067v1.

\bibitem{KKS:02}
V.~I. Kuvshinov, A.~V. Kuzmin, and R.~G. Shulyakovsky,
\newblock Acta Phys.Polon. {\bf B33}, 1721 (2002), arXiv:hep-ph/0209292.

\bibitem{KKS:03}
V.~I. Kuvshinov, A.~V. Kuzmin, and R.~G. Shulyakovsky,
\newblock Phys. Rev. E {\bf 67}, 015201 (2003), arXiv:nlin/0305028.

\bibitem{KK:05}
V.~I. Kuvshinov and A.~V. Kuzmin,
\newblock PEPAN {\bf 36}, 100 (2005).

\bibitem{Igarashi:06}
A.~Igarashi and H.~S. Yamada,
\newblock Physica D: Nonlinear Phenomena {\bf 221}, 146 (2006),
  arXiv:cond-mat/0508483.

\bibitem{Jirari:01}
H.~Jirari, H.~Kroger, X.~Q. Luo, K.~J.~M. Moriarty, and S.~G. Rubin,
\newblock Phys. Lett. A {\bf 281}, 1 (2001), arXiv:quant-ph/9910116v3.

\bibitem{Paradis:05}
F.~Paradis, H.~Kroger, G.~Melkonyan, and K.~J.~M. Moriarty,
\newblock Phys. Rev. A {\bf 71}, 022106 (2005), arXiv:quant-ph/0408051v2.

\bibitem{Brodier:01}
O.~Brodier, P.~Schlagheck, and D.~Ullmo,
\newblock Phys. Rev. Lett. {\bf 87}, 064101 (2001), arXiv:nlin/0104010v2.

\bibitem{Backer:08}
A.~B\"{a}cker, R.~Ketzmerick, S.~L\"{o}ck, and L.~Schilling,
\newblock Phys. Rev. Lett. {\bf 100}, 104101 (2008), arXiv:0707.0217v2
  (nlin.CD).

\bibitem{Polyakov:77}
A.~M. Polyakov,
\newblock Nucl. Phys. B {\bf 120}, 429 (1977).

\bibitem{Vainshtein:82}
A.~I. Vainshtein, V.~I. Zakharov, V.~A. Novikov, and M.~A. Shifman,
\newblock Sov. Phys. Usp. {\bf 25}, 195 (1982).

\bibitem{Zinn-Justin:81}
J.~Zinn-Justin,
\newblock J. Math. Phys. {\bf 22}, 511 (1981).

\bibitem{Merzbacher}
E.~Merzbacher,
\newblock {\em Quantum Mechanics} (John Wiley \& Sons, New York, 1970).

\bibitem{Skinner:88}
J.~Skinner and H.~Trommsdorff,
\newblock Journal of Chemical Physics {\bf 89}, 897 (1988).

\bibitem{Oppenlander:89}
A.~Oppenl\"ander, C.~Rambaud, H.~P. Trommsdorff, and J.-C. Vial,
\newblock Phys. Rev. Lett. {\bf 63}, 1432 (1989).

\bibitem{Rouse:95}
R.~Rouse, S.~Han, and J.~E. Lukens,
\newblock Phys. Rev. Lett. {\bf 75}, 1614 (1995).

\bibitem{Friedman:96}
J.~R. Friedman, M.~P. Sarachik, J.~Tejada, and R.~Ziolo,
\newblock Phys. Rev. Lett. {\bf 76}, 3830 (1996).

\bibitem{Friedman:00}
J.~R. Friedman, V.~Patel, W.~Chen, S.~K. Tolpygo, and J.~E. Lukens,
\newblock Nature {\bf 406}, 43  (2000).

\bibitem{Awschalom:92}
D.~D. Awschalom, D.~P. DiVincenzo, and J.~F. Smyth,
\newblock Science {\bf 258}, 414 (1992).

\bibitem{Barco:99}
E.~del Barco {\em et~al.},
\newblock Europhys. Lett. {\bf 47}, 722 (1999), arXiv:cond-mat/9810261v2.

\bibitem{Sadgrove:07}
M.~Sadgrove, M.~Horikoshi, T.~Sekimura, and K.~Nakagawa,
\newblock Phys. Rev. Lett. {\bf 99}, 043002 (2007), arXiv:0706.1627v1
  (quant-ph).

\bibitem{Dana:08}
I.~Dana, V.~Ramareddy, I.~Talukdar, and G.~S. Summy,
\newblock Phys. Rev. Lett. {\bf 100}, 024103 (2008), arXiv:0706.0871v3
  (physics.atom-ph).

\bibitem{KK:03}
V.~I. Kuvshinov and A.~V. Kuzmin,
\newblock Progress of Theor. Phys. Suppl. , 363 (2003), arXiv:nlin.CD/0305029.

\bibitem{Korn}
G.~A. Korn and T.~M. Korn,
\newblock {\em Mathematical Handbook for Scientists and Engineers: Definitions,
  Theorems, and Formulas for Reference and Review} (McGraw-Hill Book Company,
  1968).

\bibitem{Wohler:94}
C.~F. Wohler and E.~Shuryak,
\newblock Phys. Lett. B {\bf 333}, 467 (1994), arXiv:hep-ph/9402287v1.

\end{thebibliography}
\end{document}